\documentclass{article}
\usepackage{waspaa19,amsmath,graphicx,url,times}
\usepackage{color}
\usepackage{multicol}
\usepackage{multirow}
\usepackage{booktabs}
\usepackage{enumerate}   

\usepackage[utf8]{inputenc}

\title{Evaluation of post-processing algorithms for polyphonic sound event detection}

\name{L\'eo Cances, 
      Patrice Guyot, 
      Thomas Pellegrini 
      }
\address{IRIT, Universit\'e Paul Sabatier, CNRS, Toulouse, France \\ 
\{leo.cances, patrice.guyot, thomas.pellegrini\}@irit.fr 
}

\begin{document}

\ninept
\maketitle

\begin{sloppy}

\begin{abstract}

Sound event detection (SED) aims at identifying audio events (audio tagging task) in recordings, and then locating them temporally (localization task). 
This last task ends with the segmentation of the frame-level class predictions, that determines the onsets and offsets of the audio events. Yet, this step is often overlooked in scientific publications.
In this paper, we focus on the post-processing algorithms used to identify the audio event boundaries.
Different post-processing steps are investigated, through smoothing, thresholding, and optimization.
In particular, we evaluate different approaches for temporal segmentation, namely statistic-based and parametric methods.
Experiments are carried out on the DCASE 2018 challenge task 4 data. We compared post-processing algorithms on the temporal prediction curves of two models: one based on the challenge's baseline and a Multiple Instance Learning (MIL) model. Results show the crucial impact of the post-processing methods on the final detection score. Statistic-based methods yield a 22.9\% event-based F-score on the evaluation set with our MIL model. Moreover, the best results were obtained using class-dependent parametric methods with a 32.0\% F-score. 


\end{abstract}

\begin{keywords}
Weakly-labeled Sound Event Detection, Neural networks, Threshold, Post-processing
\end{keywords}

\section{Introduction}
\label{sec:intro}

In real life, sound events are produced by many possible different sources that overlap and produce a mixture.
In that context, polyphonic Sound Event Detection (SED) refers to the task of detecting overlapping audio events from a defined set of events~\cite{cakir2015polyphonic}.
This task has been investigated in various works~\cite{7472917,cakir2015polyphonic,Xia2017,kong2016deep}
and different kinds of applications that include multimedia indexing~\cite{Ballan2009soccer}, context recognition~\cite{barchiesi2015acoustic} and surveillance~\cite{harma2005automatic}.

In that domain, as well as in many others, Deep Learning~\cite{lecun2015deep} has
become a reference with deep neural networks that outperform previously proposed models~\cite{virtanen2018computational}. As these models strongly rely on data availability, the size of the exploitable corpora is expanding rapidly. The release of Audioset~\cite{audioset} is a milestone in polyphonic SED, as it provides about 5,000 hours of authentic audio recordings. Precise manual labeling of all the sound events included in this dataset is almost impossible to obtain. Therefore, Audioset is annotated only globally with a set of tags at clip-level, and the time boundaries of the audio events remain unknown. In that respect, many recent works cited here-above address the issue of semi/non-supervised SED. These works aim to find temporal sound events from learning sets annotated globally with the so-called ”weak labels.” The present study is conducted within this framework.

Typically, systems output probabilities for each event at the acoustic frame level. These temporal probabilities need to be post-processed in order to locate event onsets and offsets. In monophonic SED, the event type with the highest probability is detected as the final active event. Yet, in polyphonic SED, a threshold is often used to determine if the acoustic events are active or not~\cite{Xia2017}.
However, these post-processing methods remain globally overlooked and not described in details, as many papers focus on model descriptions.


In this paper, we evaluate different approaches for post-processing through smoothing, thresholding, and optimization.
This work aims to i) demonstrate the impact of the post-processing step on the final results, ii) document different post-processing and optimization methods (with an available implementation of code$^1$ that hopefully will benefit to the research community), iii) determine what are the best post-processing approaches for semi-supervised SED.
For this purpose, experiments are based on two different systems evaluated in the DCASE 2018 challenge.


This paper is organized as follows. Section \ref{sec:motivation} presents the semi-supervised SED task and its related works. Section~\ref{sec:method} describes post-processing approaches. We report the experimental setup in Section~\ref{sec:experiment} and analyze the results in section~\ref{sec:result}.

\section{Problem statement}
\label{sec:motivation}

\subsection{Overview}

Many recent works on semi-supervised polyphonic SED rely on the same workflow, presented in Figure~\ref{fig:workflow}.
A time/frequency image extracted from each audio file is used as input for two neural networks, a classifier, and a localizer.
The classifier outputs binary vectors representing the classes of the audio events detected in a file, namely audio tags. The localizer outputs a matrix containing the probability values for each class and each temporal frame. A segmentation algorithm is used on these probabilities to output the audio event temporal markers.

\begin{figure}[ht]
  \centering
  \centerline{\includegraphics[width=0.8\linewidth]{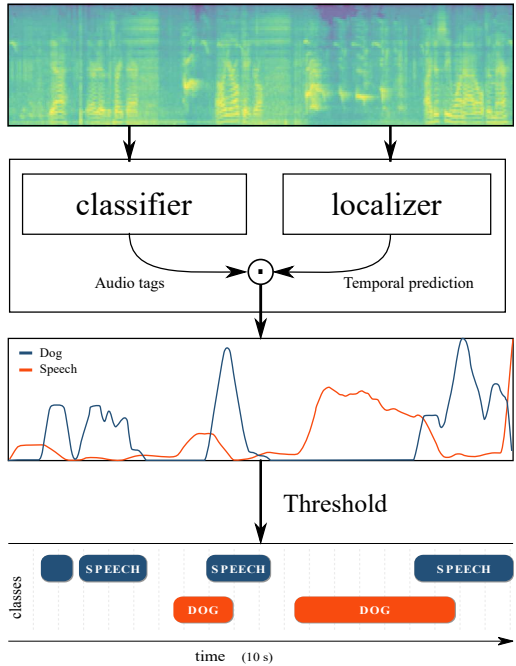}}
  \caption{Semi-supervised polyphonic sound event detection workflow illustrated for dog and speech event types as examples.}
  \label{fig:workflow}
\end{figure}

\subsection{Related work}



In different works~\cite{7472917, Guo2018}, the authors do not mention what post-processing methods are used. In~\cite{cakir2015polyphonic}, the authors report tests of eight thresholds varying from 0.1 to 0.9.
Lately, a mean-teacher model based on Recurrent Neural Networks (RNN)~\cite{Lu2018} won the DCASE 2018 challenge on large-scale weakly-labeled SED~\cite{serizel2018_DCASE}. Nevertheless, very few details are given on the post-processing process.
In~\cite{Koutini2018}, Convolutional RNNs were used to make predictions of pseudo-strong labels using median/Gaussian filters. These filters are mentioned but not fully described. In~\cite{Harb2018}, more details are given regarding which parameters must be tuned and how. The authors tested only absolute thresholding and median filtering. In other cases, simple threshold values are tuned on a development subset of the training data, such as in~\cite{Hou2018,Liu2018}.

In~\cite{Xia2017}, Xia et al. addressed the issue of threshold selection in the context of polyphonic SED. The benchmark system is a Deep Neural Network (DNN) based on~\cite{kong2016deep} and trained with binary cross-entropy as loss function. To estimate thresholds for the post-processing step, the authors proposed two approaches, named contour-based and regressor-based methods, that estimate a threshold value for each frame. In the first one, the threshold is computed as the product between a coefficient $\alpha$, which is set globally and expresses the ratio of non-empty frames, and the maximum of the probabilities of each class. The second one uses a regression to estimate the thresholds, based on an RNN, given as input the acoustic features and as a target, the probabilities outputted by the DNN. Both approaches rely on a precise labeled training set that contains the time boundaries of each audio event. 

Finally, aside from this last work, post-processing methods are often overlooked and not carefully evaluated. To our knowledge, there is no systematic analysis of the impact of post-processing within the sound event detection task. We suppose that many research works could benefit from a clear presentation of the approaches as well as a detailed evaluation.

\section{Post-processing methods}
\label{sec:method}






This section presents the post-processing approaches that are evaluated, namely smoothing, segmenting and optimizing.

Smoothing removes noise in the probabilities, limiting the number of small segments or small gaps that will be created during the segmentation process. In our work, we use a smoothed moving average.
The smoothing of the temporal prediction output by the model can be class-dependent as the smoothing window size may change with the class.

    

\subsection{Segmentation}

\subsubsection{statistic-based methods}
The statistic-based methods are directly based on the statistics extracted from the temporal predictions of each sample. The main advantage of these methods is that they are fast and often efficient. We apply them on the test and evaluation sets. We refer to them as: 
i) class-independent data-wise average (CIDWA); ii) class-dependent data-wise average (CDDWA).
We also tested class-(in)dependent file-wise average and median. However, those methods will not be mentioned as they yield either poor results (file-wise average/median) or slightly worse results (data-wise median).

\begin{enumerate}[(i)]
    
    \item CIDWA: we use the localizer outputs to compute the average probability of each class over time. We aggregate the averages to create a single threshold for all the classes.
    
    \item CDDWA: for this variant, class-dependent averages are used as thresholds.
    
\end{enumerate}

\subsubsection{Parametric methods}
The parametric methods require optimization. We optimized the parameters on the test set and used them on the evaluation set. They can either be class-independent or class-dependent. These methods are three in number and are called i) class-(in)dependent absolute (CIA-CDA), ii) class-(in)dependent hysteresis (CIH - CDH), iii) class-(in)dependent slope (CIS - CDS).

\begin{enumerate}[(i)]
    \item Absolute thresholding refers to directly applying a unique and arbitrary threshold to the temporal predictions without using their statistics. This naïve approach still yields exploitable results that can get close to the best ones in some cases. It is also the approach with the shortest optimization time due to the unique parameter to optimize.  
    
    \item Hysteresis thresholding consists of two thresholds. One of them will be used to determine the onset of an event, and the second one its offset. This algorithm is used when probabilities are unstable and changing at a high pace. It should, therefore, decrease the number of events detected by the algorithm and reduce the insertion and deletion rates, giving a better error rate than the Absolute threshold approach.
    
    \item The Slope-based method uses a different approach that does not use any fixed threshold. Its principle is to determine the start and end of a segment by detecting fast changes in the probabilities over time. Fast-rising probabilities will signify the start of a segment, and fast decreasing probabilities or a plateau, its end. It is capable of detecting the end of segments even if the probabilities are high. However, it is very dependent on the model's outputs.
\end{enumerate}

\subsection{Optimization}
\label{sec:opti}

The parametric methods regroup together the different algorithms that exploit arbitrary parameters to locate with precision sound events. The search for the best parameter combination is a meticulous work that is often not possible to automatize. Indeed, depending on the number of parameters to tune, the search space growth is exponential and the execution time often exceeds reasonable times. Consequently, we implemented a smarter exploration method called dichotomic search.

For every parameter to tune, the user provides global boundaries and, in between these boundaries, the algorithm tries every combination with a coarse resolution and picks the one that yields the best score. From this combination, new -- smaller -- boundaries are computed. The complete process is then repeated in between the new limits, with every step increasing the precision of each parameter and reducing the search space. It stops when the number of steps given by the user is reached.

The dichotomous search algorithm,  when compared to an exhaustive search of all the possible combinations, considerably reduces the time needed to reach a near-optimal solution with excellent accuracy. However, the execution time is still dependent on the number of parameters to tune and the amount of iterations for every step. The total number of combinations increases exponentially.



\section{Experiments}
\label{sec:experiment}

\subsection{Audio Material}

The DCASE 2018 challenge task 4~\cite{dcase2018web} provided audio material directly extracted from Audioset. The training set is divided into three subsets. Only one of them is weakly annotated and we will refer to it as the "weak" subset. The two others, being not annotated at all, are not of any use for the training of our models. The weak training subset is comprised of 1578 clips (2244 class occurrences) for which weak annotations have been verified and cross-checked.

Along with these three subsets, the challenge also proposed a test and an evaluation subset. Both of them have been strongly annotated, providing precise temporal localization (onset and offset boundaries) for each event occurrence and are composed respectively of 279 and 880 files. Both of them present a similar distribution of the classes.

Each file can include one or several events from a set of  sound classes occurring in domestic environments: \emph{Speech}, \emph{Dog}, \emph{Cat}, \emph{Alarm/ Bell ringing}, \emph{Dishes}, \emph{Frying}, \emph{Blender}, \emph{Running water}, \emph{Vacuum cleaner}, and \emph{Electric shaver/toothbrush}.
All the files are 10-second clips extracted from Audioset.
These recordings contain generally several overlapping audio events from different classes.

The parametric methods will be optimized using the test dataset and validated on the evaluation dataset.



\begin{figure}[ht]
  \centering
  \centerline{\includegraphics[width=0.8\linewidth]{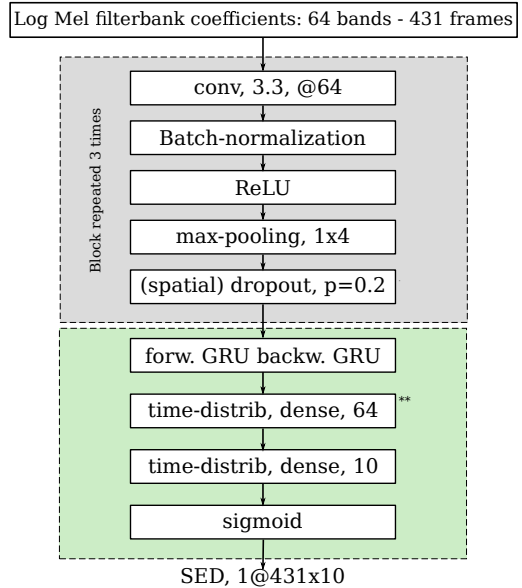}}
  \caption{Architecture of the models MIL and Baseline. In the baseline model, ($*$) is replaced by a dropout layer ($p=0.3$), and ($**$) is removed.}
  \label{fig:mil_sed}
\end{figure}

\subsection{Models}
\label{models}

To observe the impact of the segmentation algorithms, we use two models. Both models output a matrix of size 431 x 10 representing the prediction score for each of the ten classes over time. 
The first model is based on "Multiple Instance Learning (MIL)" as proposed in~\cite{salamon2017dcase,Cances2018,Pellegrini2019cosine}. It consists of two networks: one for classification and a second one for temporal prediction.
Their architecture is described in Figure~\ref{fig:mil_sed}.

The second model is similar to the DCASE 2018 task 4 challenge baseline~\cite{serizel:hal-01850270}, that will be referred to as Baseline hereafter.
It is composed of a convolutional part followed by a recurrent one, namely a bi-directional Gate Recurrent Unit layer and a time-distributed dense layer.

\subsection{Experimental setup}

\begin{table*}[t]
    \centering
\begin{tabular}{l|llll|llll|c}
\toprule
\multirow{3}{*}{\textbf{Post-processing methods}}              & \multicolumn{4}{c|}{\textbf{Baseline}}                                 & \multicolumn{4}{c|}{\textbf{MIL}}                                       & Relative \\ 
                                               & \multicolumn{2}{c}{\textbf{Test}} & \multicolumn{2}{c|}{\textbf{Eval}} & \multicolumn{2}{c}{\textbf{Test}} & \multicolumn{2}{c|}{\textbf{Eval}}  & Computation \\
                                               & F1 (\%)          & Er             & F1 (\%)          & Er              & F1 (\%)          & Er              & F1 (\%)          & Er             & Time \\ \midrule
\textit{Coarse grid search}                    & 20.9             & 1.2            & \textit{19.4}    & \textit{1.3}   & \textit{18.2 }   & \textit{1.8 }   & \textit{15.3 }   & \textit{1.8 }   & \hphantom{111}1 \\ \midrule
Class-independent data-wise average (CIDWA)    & 19.9             & 1.3            & 17.9             & 1.5             & 29.8             & 2.0             & 25.8             & 2.5            & \hphantom{111}0 \\
Class-dependent data-wise average (CDDWA)      & 19.6             & 1.3            & \textbf{18.7}    & 1.4             & 32.5             & 1.8             & \textbf{29.9}    & 2.4            & \hphantom{111}0 \\ \midrule
Class-independent absolute (CIA)               & 25.0             & 1.1            & \textbf{22.8}    & 1.2             & 44.2             & 1.1             & 37.1             & 1.4            & \hphantom{111}1 \\
Class-independent Hysteresis (CIH)             & 25.0             & 1.1            & 22.6             & 1.2             & 46.4             & 1.0             & \textbf{40.7 }   & 1.2            & \hphantom{111}3 \\
Class-independent Slope (CIS)                  & 24.3             & 1.2            & 21.0             & 1.2             & 43.6             & 1.2             &  35.5            & 1.5            & \hphantom{1}115\\ \midrule
Class-dependent absolute (CDA)                 & 26.5             & 1.1            & 22.3             & 1.3             & 53.2             & 0.9             & \textbf{43.9}    & 1.2            & \hphantom{11}10 \\
Class-dependent Hysteresis (CDH)              & 26.5              & 1.1            & 23.0             & 1.2             & 53.1             & 0.8             & 42.9             & 1.1            & \hphantom{11}29\\
Class-dependent Slope (CDS)                    & 26.2             & 1.1            & \textbf{23.4}    & 1.2             & 52.4             & 0.9             & 41.0             & 1.2            & 1155 \\
\bottomrule
\end{tabular}
\caption{F1-score and Error Rate for both baseline and MIL models on test and evaluation set with the AT Oracle. The last column shows the relative computation time of each method compared to CIA.\label{tab:results}}
\end{table*}

Post-processing takes place after the model training phase, when thresholds are applied on smoothed time predictions to obtain the onset and offset of audio events. It is performed in the following order: 1) The curves representing the prediction of the model for each frame are smoothed using the smoothed moving average algorithm. This smoothing was applied only with the parametric methods since statistic-based methods are not meant to involve optimization. 
2) the temporal predictions are segmented using one of the segmentation algorithms described above.
3) Segments separated by a gap smaller than the challenge tolerance margin are merged together. In the same fashion, segments smaller than this margin are removed. 

When a parametric method is used, the process is repeated to reach the best score by using the optimization algorithm. Similarly, the smoothing window size can be optimized either class-independently or class-dependently.

For both models, we tested the localization algorithms previously described, as well as a coarse grid search that represents the combination of absolute thresholds from 0.1 to 0.9 and a 0.1 step, with smoothing window sizes from 5 to 21 and a step of 2, totaling 64 combinations. 
Two issues must be taken into consideration: i) the potential errors made by the audio tagging model, ii) the setting of parameter values when using a parametric method.




\begin{enumerate}[(i)]
    \item To remove the bias induced by faulty audio tag classification, we used the classes of the strong annotations as if they were outputs from a perfect classifier. It allows us to pick only the relevant classes on which the events must be localized. We will refer to this mode as \emph{Audio Tagging oracle} (AT oracle). We applied this procedure on both test and evaluation subsets.
    
    \item We used the event-based metrics defined in~\cite{Mesaros2016}. More precisely the macro-F1 score, alias F1, with the challenge precision parameters: a 200ms collar on the onsets and an offset collar corresponding to 20\% of the event's length.
    
\end{enumerate}

\section{Results}
\label{sec:result}
The results are presented in Table 1. Overall, they show a wide disparity in values. The F1 score varies from 17.9\% to 23.4\% with our baseline model, and from 25.8\% to 43.9\% with the MIL model. Therefore, we observe a significant impact of the post-processing algorithms on the final results. The best scores are obtained by using the class-dependent parametric methods. On the evaluation set, CDS for our baseline gives a final F1 score of 23.4\%, and CDA for the MIL model a final F1 score of 43.9\%.

Regarding computation time, the best method is not necessarily the longest one and the gain, if there is any, is not linear. The baseline benefits only of 1.1 points for a computation time a thousand time longer, whereas MIL shows a decrease in performance. However, the gain from class-independent to class-dependent is worth the extra execution time, which is in our case, approximately ten times more.

With a closer look at statistic-based methods, CDDWA yields better performance with a final F1 score of 18.7\% and 29.9\% respectively on our baseline and MIL. In both cases, it gave better results on the evaluation subset than CIDWA. The Class-Dependent variant of the algorithm seems more suitable than the Class-Independent one even though it gave a slightly worse result on the test set (0.3 absolute difference). Indeed, If we look closely at the transition between test and evaluation sets, the difference is only of -4\% for the class-dependent and -10\% relative for the class-independent, making the first more robust.

Ultimately, parametric methods present the best results. All methods perform better than the manually chosen threshold or the statistic-based ones. Furthermore, the best scores are obtained using their class-dependent variant. The same observation can be done between test and evaluation set as the difference is only of -8\% for the class-dependent, and -18\% for the class-independent, making the class-dependent method not only perform better but also more robust. Our baseline reachS, on the evaluation set, a final F1 score of 23.4\% with CDS, it represents an improvement of 4 points (or 20.6\%). For MIL, it is the CDA method that yields the best final F1 score with an improvement of 28.6 points (or 187.0\% relative).

If the statistic-based methods have already shown improvement of the final F1 score, the Parametric pushes it even further, especially the class-dependent variant of the algorithms. The maximum F1 score for the statistic-based methods is 18.7\% and 29.9\% for the baseline and the MIL respectively when for the parametric one they are 23.4\% and 43.9\%.

Regarding smoothing, in the class-dependent parametric methods, the smoothing window size is a parameter that can be optimized. A closer look at the parameter combination resulting from the optimization shows a wide variety of window size from 9 (\emph{Dishes}) to 27 (\emph{Vacuum cleaner}) frames. This highlights the importance of smoothing the predictions according to the classes.

When looking at the scoring of each class independently, the improvement is uniformly dispatched. When optimizing the algorithm parameters specifically for each class, (class-dependent parametric methods), almost every class seem to benefit from the optimization, but few do not. It is the case with the class \emph{dishes} for instance. Moreover, the algorithm tested does not seem to show the clustering of some class together.

Finally, we applied these methods on our model without using the AT Oracle but the audio tags outputted by their classifier. We then compared it to the best models from the DCASE 2018 task 4 challenge. After optimization, the baseline F1 score increases from 12.6\% to 14.1\%, and for MIL, from 21.1\% to 32.0\%. The first~\cite{Lu2018} and the second~\cite{Liu2018} ranked participants obtained a F1 score of 32.4\% and 29.9\%, respectively.




\section{Conclusion}
In SED, prediction post-processing is often overlooked. There is no systematic analysis of its impact, and we compare several solutions to this problem. We explored several methods to segment the temporal prediction outputs from DNN-based models that can be divided into two categories: statistic-based and parametric approaches, either class-independent or class-dependent.

The methods presented show the impact post-processing can have on the final performance. Statistic-based methods do not require optimization, making them suitable for a quick preview of the results that can be achieved. They are model-agnostic, easy to implement, fast to compute and can produce better results than a coarse grid search of the smoothing and thresholding parameters.

The parametric methods are nonetheless better. Our best model shows an improvement of 28.6 points (187.0\% relative) by using the class-dependent absolute method. The class-dependent methods do not only yield greater results but also greater robustness when switching from the test set to the evaluation set.

When it comes to the numerous datasets available nowadays, a larger one could be used to scale these methods and confirm their relevance. The same applies to the vast variety of models that have been implemented for SED tasks. Furthermore, other optimization techniques, relying on genetic algorithms, for instance, could be added to the ones tested in this work.

\newpage

\bibliographystyle{IEEEtran}
\bibliography{refs}
%
%
%
%
%
%
%
%
%

\end{sloppy}
\end{document}